\documentclass{article}
\usepackage{graphicx}
\usepackage{amsmath}   
\usepackage{amssymb}
\usepackage{epstopdf}
\usepackage{authblk}
\usepackage{multirow}
\usepackage{caption,subcaption}
\begin{document}
\title{Exact dissipative solutions and gravitational complexity of evolving shear-free radiating stellar envelopes.}

\author{$^*$Vusi M Zitha\\ Department of Mathematics, Faculty of Applied Sciences, Durban University of Technology, Durban 4000, South Africa. 
		\\vusiz@dut.ac.za\\
  Megandhren Govender\\ Department of Mathematics, Faculty of Applied Sciences, Durban University of Technology, Durban 4000, South Africa. \\megandhreng@dut.ac.za }



\date{}
\maketitle
\begin{abstract}
	We study the gravitational behaviour of a shear-free radiating relativistic star undergoing dissipative evolution by matching a spherically symmetric interior spacetime with heat flux to the exterior Vaidya spacetime. Through suitable transformations, the nonlinear junction condition is simplified, leading to two new classes of exact solutions and the construction of an explicit radiating stellar model with closed-form gravitational potentials. The physical properties of the model are examined through the evolution of the energy density, radial pressure, tangential pressure, and heat flux, while the gravitational complexity is analysed using Herrera's complexity factor, which separates the effects of pressure anisotropy, density inhomogeneity and dissipative heat transport. The results show that the dissipative heat-flux contribution can become the dominant component of gravitational complexity near the outer boundary of the stellar envelope. Our results provide new insights into the role of dissipation in the dynamical evolution and gravitational complexity of self-gravitating radiating systems.
\end{abstract}
\section{Introduction}

The dynamical evolution of self-gravitating stellar systems provides an important framework for studying the interplay between gravity, matter, radiation and dissipative processes. The pioneering work of Oppenheimer and Snyder~\cite{OppenheimerSnyder1939} established a foundation for the relativistic description of evolving stellar configurations. In realistic astrophysical systems, however, the evolution of compact objects is accompanied by processes such as heat transport and radiation. Relativistic models incorporating dissipative effects are therefore essential for understanding the dynamical behaviour and gravitational structure of radiating stellar matter.

A radiating spherical configuration is commonly described by matching an interior spacetime containing a dissipative fluid to an exterior Vaidya spacetime~\cite{Vaidya1951}. The Vaidya solution provides a suitable description of the exterior geometry of a spherically symmetric radiating source, while the junction conditions derived by Santos~\cite{Santos1985} ensure a consistent matching between the interior and exterior regions. These conditions lead to a nonlinear differential equation involving the gravitational potentials at the stellar boundary, making the construction of exact radiating solutions a challenging problem. The Misner--Sharp mass provides an invariant measure of the gravitational mass contained within a spherical region and is particularly useful in the analysis of dynamical relativistic configurations~\cite{MisnerSharp1964}. Earlier investigations have also demonstrated the importance of dissipative processes in the evolution of radiating relativistic stars~\cite{Glass1989,ThirukkaneshMoopanarGovender2012,GovenderReddyMaharaj2009,GovenderThirukkanesh2015,Ivanov2016}.

Exact solutions of the Einstein field equations are valuable because they provide analytical models in which the geometry, matter variables and dynamical properties can be studied explicitly. However, the nonlinear nature of the junction condition significantly restricts the direct construction of closed-form solutions. Various techniques have consequently been developed to simplify the governing equations, including symmetry methods, generating functions and transformations of the metric potentials. In particular, transformations involving the gravitational potentials have proved effective in reducing the complexity of the field equations and generating exact relativistic stellar models. Ngubelanga and Maharaj~\cite{SifisoandSDMaharaj} introduced a transformation of this type in the study of static stellar configurations, and subsequent investigations extended related techniques to broader classes of relativistic solutions~\cite{SifisoandSDMaharaj2,NgubelangaMaharaj20153,NgubelangaMaharajRay2015}. These results motivate the use of transformed gravitational potentials in the present dissipative setting.

An additional aspect of relativistic stellar structure is the concept of gravitational complexity. Herrera~\cite{Herrera2018} introduced a complexity factor based on the orthogonal splitting of the Riemann tensor. This formalism provides a covariant description of structural complexity and establishes a connection between density inhomogeneity and pressure anisotropy. Herrera, Di Prisco and Ospino~\cite{HerreraDiPriscoOspino2018} subsequently demonstrated how these competing physical effects can be incorporated into the complexity factor of relativistic fluids. The formalism has since been applied to a variety of anisotropic and dynamical stellar configurations.

The complexity framework was further extended to dissipative fluids by Herrera, Di Prisco and Ospino~\cite{HerreraDiPriscoOspino2020}, where the heat flux introduces an additional contribution to the gravitational complexity. Bogadi \textit{et al.}~\cite{BogadiGovenderMoyo2022a} developed a decomposition in which the complexity factor is separated into contributions associated with pressure anisotropy, density inhomogeneity and dissipative heat transport. This decomposition provides a useful means of identifying the relative influence of irreversible thermodynamic processes on the gravitational structure of a self-gravitating fluid. More recent investigations have applied this framework to dynamical and dissipative relativistic stellar models~\cite{GovenderBarmanSharma2023,GovenderBogadiGovenderMewalal2024,MajoziGovenderMaharajGovender2025,MajoziGovenderMaharaj2025a,MajoziMaharajGovender2026}.

Despite these developments, there remains scope for constructing exact dissipative radiating configurations for which the different sources of gravitational complexity can be evaluated explicitly and followed simultaneously in space and time. In particular, it is desirable to obtain analytical models in which pressure anisotropy, density inhomogeneity and dissipative heat transport can be examined within a common dynamical framework. The nonlinear junction condition remains the principal mathematical obstacle to achieving this objective.

In the present work, we investigate a shear-free, spherically symmetric radiating stellar configuration and employ suitable transformations of the gravitational potentials to simplify the nonlinear junction condition. Two classes of exact solutions are obtained. The first follows from a linear choice for the transformed radial potential, while the second is generated through a first-order reduction of the nonlinear junction equation. An explicit member of the second class is then selected for detailed physical investigation. The resulting metric potentials and matter variables are obtained in closed form, allowing the dynamical behaviour of the radiating configuration to be examined analytically and numerically.

The exact solution obtained in this work is singular at the centre and is therefore interpreted as describing a stellar envelope rather than a complete stellar interior. Core--envelope descriptions provide a useful framework for relativistic stellar modelling when an exact solution is not regular at the centre. Takisa \textit{et al.}~\cite{TakisaMaharajMulangu2019} investigated relativistic configurations with distinct core and envelope regions, while Khunt \textit{et al.}~\cite{KhuntThomasVinodkumar2023} further demonstrated the usefulness of the core--envelope approach in the construction and analysis of relativistic stellar models. Accordingly, the physical analysis in the present work is restricted to the finite envelope domain
\[
r_0\leq r\leq r_{\Sigma},\qquad r_0>0,
\]
where ($r_0$) denotes the inner boundary of the envelope and ($r_{\Sigma}$) the stellar surface.

A principal objective of the present study is to examine the gravitational complexity of the resulting dissipative envelope. The complexity factor is decomposed into three contributions,
\[
Y_{TF}=Z_1+Z_2+Z_3,
\]
representing, respectively, pressure anisotropy, density inhomogeneity and dissipative heat transport. This decomposition enables the relative importance of these physical mechanisms to be investigated during the evolution of the radiating configuration. In contrast to analyses restricted primarily to radial profiles, we consider the simultaneous dependence of the matter variables and complexity factor on both the radial coordinate and time. The mass function is also examined as an additional measure of the evolving gravitational content of the configuration. This provides a unified description of the geometry, matter distribution, dissipative transport and gravitational complexity of the exact radiating envelope~\cite{ZithaGovenderGovender2022}.

The paper is organised as follows. Section~\ref{Stellar} presents the Einstein field equations for a shear-free radiating fluid and introduces the transformed gravitational potentials. Section~\ref{Expansion} discusses the expansion and temporal evolution of the configuration. Section~\ref{Exterior} introduces the exterior Vaidya spacetime and the relevant matching conditions, while Section~\ref{junction} derives the resulting nonlinear junction equation. Two classes of exact solutions are obtained in Sections~\ref{Linear} and~\ref{ODEP}, respectively. Section~\ref{complexit} introduces the gravitational complexity formalism. Section~\ref{Analysis} investigates the physical properties of the selected exact solution, including the matter variables, expansion scalar, mass function and the different contributions to the complexity factor. Finally, Section~\ref{Conclusion} summarises the main results and outlines directions for future work.

\section{Stellar characteristics\label{Stellar}}

The line element for the interior geometry of a radiating star for a
general spherically symmetric shear-free, spacetime can be written as
\begin{equation}
	ds^2 = -A^2(r,t)dt^2 + B^2(r,t)[dr^2+r^2(d\theta^2 +
	\sin^2{\theta}d\phi^2)] \, ,\label{metric}
\end{equation}
where $A = A(t,r)$ and $B = B(t,r)$ are metric functions.The radial unit four-vector $\chi^a$, orthogonal to the fluid
four-velocity $u^a$, is defined by
\begin{equation}
	\chi^a=\frac{1}{B}\delta^a_1
	=(0,B^{-1},0,0),
	\qquad
	\chi_a=(0,B,0,0),
\end{equation}
such that
\begin{equation}
	u_a\chi^a=0,
	\qquad
	\chi_a\chi^a=1.
\end{equation}
The heat-flow vector is taken to be directed along the radial
unit vector, so that
\begin{equation}
	q^a=q\chi^a,
	\qquad
	q_a=q\chi_a.
\end{equation}
The energy-momentum tensor for the interior matter distribution
is then given by
\begin{equation}
	T_{ab}
	=(\mu+P_\perp)u_a u_b
	+P_\perp g_{ab}
	+(P_r-P_\perp)\chi_a\chi_b
	+q_a u_b+q_bu_a.\label{momentum}
\end{equation}
where $\mu$ is the energy density, $P_r$ is the radial pressure,
$P_{\perp}$ is the tangential pressure, $q_a$ is the heat-flow vector.

The Einstein field equations with coupling constant taken to be unity
\begin{equation}
	G_{ab} =T_{ab}, \label{cosmological}
\end{equation}
for the metric (\ref{metric}) and the matter distribution (\ref{momentum}), can be written as
\begin{eqnarray}
	\mu  &=& \frac{3}{A^2}\left(\frac{\dot{B}}{B}\right)^2 - \frac{1}{B^2}\left[2\frac{B''}{B} + \left(\frac{B'}{B}\right)^2 - \frac{4}{r}\frac{B'}{B} \right]  ,  \label{14a} \\ \nonumber \\
	P_r  &=&-\frac{1}{A^2}\left [2\frac{{\ddot B}}{B}-\left(2\frac{\dot A}{A}-\frac{\dot B}{B}\right)\frac{\dot B}{B} \right] \nonumber \\
	&& +\frac{1}{B^2}\left(2\frac{A'}{A}+\frac{B'}{B}+\frac{2}{r}\right)\frac{B'}{B}+\frac{2}{r}\frac{A'}{A}\frac{1}{B^2}, \label{14b}  \\ \nonumber \\
	P_{\perp} &=& -\frac{1}{A^2}\left [2\frac{{\ddot B}}{B}-\left(2\frac{\dot A}{A}-\frac{\dot B}{B}\right)\frac{\dot B}{B} \right]\nonumber \\&&+  \frac{1}{B^2}\left[\frac{A''}{A} + \frac{B''}{B} +\frac{1}{r} \frac{A'}{A}+\frac{1}{r}\frac{B'}{B} +\left(\frac{B'}{B}\right)^2\right], \label{14c}\\ \nonumber \\
	q  &=& \frac{2}{AB^2}\left(\frac{\dot{B'}}{B} - \frac{\dot
		B}{B}\frac{B'}{B} - \frac{\dot B}{B}\frac{A'}{A}\right),
	\label{14d}
\end{eqnarray}

where the dot and prime denote derivatives with respect to time and radial distance, respectively.

We consider a comoving fluid distribution with four-velocity
\begin{equation}
	u^a=\frac{1}{A}\delta^a_0,
	\qquad
	u_a=(-A,0,0,0).
\end{equation}
The corresponding kinematical quantities associated with the metric (\ref{metric}) are given by
\begin{eqnarray}
	\dot{u}_a&=&\left(0,\frac{A'}{A},0,0\right),\\ \nonumber\\
	\Theta&=&\frac{3\dot{B}}{AB} \label{Theta},\\ \nonumber \\
	\sigma_{ab}&=&0,\\ \nonumber \\
	\omega_{ab}&=&0,
\end{eqnarray}
where $\dot{u}_a$, $\Theta$, $\sigma_{ab}$, and $\omega_{ab}$ denote the four-acceleration, expansion scalar, shear tensor, and vorticity tensor, respectively. The vanishing of the shear and vorticity tensors confirms that the fluid flow is shear-free and irrotational.

We obtain the pressure-isotropy condition from Eqs. (\ref{14b}) and (\ref{14c}).

\begin{equation}
	\left[ \frac{A''}{A}+\frac{B''}{B}-\left(2\frac{B'}{B} +\frac{1}{r}\right)\left(\frac{A'}{A}+\frac{B'}{B}\right) \right]=0 \label{pressure}
\end{equation}

We introduce the transformation

\begin{equation}
	x = r^2, \hspace{1cm} L =B^{-1}, \hspace{1cm} G = L A,
\end{equation}
where $L$ and $G$ are functions of $x$ and $t$. This transformation was successfully employed by Ngubelanga and Maharaj to construct exact models of static stars.
The field equations (\ref{14a})-(\ref{14d}) can be recast as

\begin{eqnarray}
	\mu&=&\frac{3l^2_t}{G^2}+4\left[2xL_xLL_{xx}-3\left(xL_x-L \right)L_x\right],\\ \nonumber\\
	P_r&=&\frac{1}{G^2}\left[2L_{tt}L-3L^2_t-2\frac{G_t}{G}L_tL\right]\nonumber \\&&+4L\left(L-2xL_x\right)\frac{G_x}{G}-4\left(2L-3xL_x\right)L_x,\\ \nonumber \\
	P_{\perp}&=&\frac{1}{G^2}\left[2L_{tt}L-3L^2_t-2\frac{G_t}{G}L_tL\right]\nonumber\\&&+4xL^2\frac{G_{xx}}{G}+4L\left(L-2xL_x\right)\frac{G_x}{G}\nonumber \\ &&-4\left(2L-3xL_x\right)L_x-8xLL_{xx},\\ \nonumber \\
	q&=&\frac{4\sqrt{x}L^2}{G}\left(-L_{xt}+\frac{G_x}{G}L_t\right)	
\end{eqnarray}
The above set of field equations generalise the static formalism of Ngubelanga and Maharaj to the dissipative regime \cite{SifisoandSDMaharaj}.

\section{Expansion and Temporal Evolution\label{Expansion}}

Using the transformations introduced above, the pressure isotropy condition given by equation~(\ref{pressure}) reduces to
\begin{equation}
	LG_{xx}+2G_xL_x-2GL_{xx}=0,
	\label{pressure1}
\end{equation}
where subscripts denote partial differentiation with respect to the corresponding variable.

The expansion scalar provides an invariant measure of the rate of change of the proper volume of a fluid element along the fluid four-velocity. For a congruence of world lines with four-velocity $u^{a}$, the quantity $\Theta=u^{a}{}_{;a}$ determines whether neighbouring fluid elements are moving apart or towards one another during the evolution of the system. Using the transformations introduced above and equation~(\ref{Theta}), the expansion scalar takes the simple form
\begin{equation}
	\Theta=-\frac{3L_t}{G}. \label{Theta1} 
\end{equation} 
Since the spacetime is shear-free and irrotational, the expansion scalar completely characterises the kinematical evolution of the fluid distribution. In particular, all changes in the local volume of the fluid occur isotropically through $\Theta$, without distortion from shear stresses or rotational effects. Consequently, the sign and magnitude of $\Theta$ play an important role in determining the temporal behaviour of the radiating stellar envelope. Assuming that the metric function $G$ remains positive throughout the evolution, the sign of $\Theta$ is governed entirely by the derivative $L_t$. The three possible evolutionary regimes are therefore

\begin{equation}
	L_t<0 \quad\Longrightarrow\quad \Theta>0,
\end{equation} corresponding to expansion, \begin{equation}
	L_t=0 \quad\Longrightarrow\quad \Theta=0, 
\end{equation} which describes a locally static configuration, and 
\begin{equation}
	L_t>0 \quad\Longrightarrow\quad \Theta<0, \end{equation}
corresponding to contraction. Since $L=B^{-1}$, these conditions may also be interpreted in terms of the metric potential $B$. During contraction ($\Theta<0$), the quantity $L$ increases with time and consequently $B$ decreases, indicating a reduction in the physical volume occupied by the fluid. Conversely, during expansion ($\Theta>0$), the quantity $L$ decreases with time and $B$ increases, corresponding to an increasing fluid volume. For a dissipative radiating configuration, the behaviour of $\Theta$ is particularly significant because it determines how the stellar envelope evolves while simultaneously losing energy through the heat flux $q$. The expansion scalar therefore provides a direct connection between the kinematical evolution of the fluid and the thermodynamical processes responsible for energy transport. In the exact solutions obtained later, the explicit form of $\Theta$ enables the dynamical state of the model to be identified and allows the temporal evolution of the radiating envelope to be monitored throughout its evolution. The relation (\ref{Theta1}) further indicates that the magnitude of $\Theta$ is controlled by the ratio $|L_t|/G$. Consequently, rapid temporal variations of the gravitational potential $L$ lead to large values of the expansion scalar and hence stronger dynamical evolution, whereas slowly varying configurations correspond to small values of $|\Theta|$. The expansion scalar and its physical interpretation for the selected exact solution are discussed in Section~\ref{Analysis}, where the temporal evolution of the radiating envelope is investigated in detail.

\section{Exterior spacetime\label{Exterior}}

The exterior spacetime of the radiating star is described by the outgoing Vaidya solution
\begin{equation}
	ds^{2}
	=
	-\left(1-\frac{2m(v)}{r}\right)dv^{2}
	+2\,dv\,dr
	+r^{2}\left(d\theta^{2}+\sin^{2}\theta\,d\phi^{2}\right),
	\label{Vaidya}
\end{equation}
where $v$ is the retarded null coordinate and $m(v)$ is the gravitational mass of the star as measured by a distant observer. The metric (\ref{Vaidya}) describes the exterior gravitational field of a radiating star surrounded by outgoing null radiation.

The smooth matching of the interior spacetime to the exterior Vaidya solution across the timelike boundary hypersurface \(\Sigma\) is achieved by imposing the junction conditions derived by Santos \cite{Santos1985}. The continuity of the first and second fundamental forms across \(\Sigma\) yields the total mass enclosed within the boundary,
\begin{equation}
	m(v)
	=
	\left[
	\frac{Y}{2}
	\left(
	1+\frac{\dot{Y}^{\,2}}{A^{2}}
	-\frac{Y'^{\,2}}{B^{2}}
	\right)
	\right]_{\Sigma},
	\label{mass}
\end{equation}
where \(Y=rB(r,t)\). Using the transformation, we obtain

\begin{equation}
	m(v)
	=
	\left[
	\frac{\sqrt{x}}{2L}
	\left(
	1
	+\frac{xL_t^{\,2}}{L^{2}G^{2}}
	-\frac{(L-2xL_x)^{2}}{L^{2}}
	\right)
	\right]_{\Sigma}.
	\label{massLG}
\end{equation}
where the subscript $\Sigma$ indicates that the quantity is evaluated at the boundary surface of the star.

\section{Junction condition\label{junction}}

The interior spacetime is matched smoothly to the exterior Vaidya spacetime across the boundary hypersurface $\Sigma$. The continuity of the first and second fundamental forms leads to the Santos junction condition.
\begin{equation}
	(P_r)_{\Sigma}=(Bq)_{\Sigma},
\end{equation}

yields the nonlinear master equation
\begin{eqnarray}
	0&=&\frac{1}{G^2}\left[2LL_{tt}-3L_t^2-2\frac{G_t}{G}LL_t\right]\nonumber\\
	&&+4L\left(L-2xL_x\right)\frac{G_x}{G}
	-4\left(2L-3xL_x\right)L_x\nonumber\\
	&&+\frac{4\sqrt{x}L}{G}\left(L_{xt}-\frac{G_x}{G}L_t\right),
	\label{master}
\end{eqnarray}
where $x$ is treated as constant.
Equation (\ref{master}) is a highly nonlinear partial differential equation and, in general, does not admit a closed-form solution. Consequently, obtaining exact solutions requires the introduction of an appropriate simplifying assumption or choices that preserves the physical structure of the model while reducing the equation to a tractable form. In the following section, we introduce such a condition, enabling the master equation to be integrated exactly.

\section{Solution under Linear Approach for $L(x,t)$\label{Linear}}
To simplify the master equation~(\ref{master}) and facilitate the construction of exact solutions, we impose the auxiliary condition
\begin{equation}
	L_{xt}-\frac{G_x}{G}L_t=0.
	\label{Sol2}
\end{equation}
which leads to 
\begin{equation}
	\left(\ln G\right)'=\frac{L_{xt}}{L_t}.\label{Sol3}
\end{equation}
To determine an explicit expression for $G$ from equation (\ref{Sol3}) and facilitate the integration of the master equation, we assume that $L$ is linear in the transformed radial coordinate $x$, namely
\begin{equation}
	L=a(t)+b(t)x,
	\label{Sol4}
\end{equation}
where $a(t)$ and $b(t)$ are arbitrary functions of time. Substituting (\ref{Sol4}) into equation (\ref{Sol3}) and integrating, we obtain
\begin{equation}
	G=k(t)\big(\dot{a}(t)+\dot{b}(t)x\big)=k(t)S,
	\label{Sol5}
\end{equation}
where $k(t)$ is an arbitrary function of time and  $S=\dot{a}(t)+\dot{b}(t)x$.
From equation (\ref{Sol5}) it follows that

\begin{equation}
	\frac{G_x}{G}=\frac{\dot{b}}{S}\hspace{1cm} \frac{G_t}{G}=\frac{\dot{k}}{k}+\frac{S_t}{S} \label{Sol6}.
\end{equation}
Substituting equation (\ref{Sol6}) and (\ref{Sol4}) into (\ref{master}), we obtain
\begin{eqnarray}
	0&=&-\frac{1}{k^2}\left(3+2\frac{\dot{k}}{k}\frac{L}{S}\right)+4\frac{\dot{b}}{S}L\left(a-bx\right)\nonumber\\&&+\left(-8ab+4b^2x\right) \label{Sol7}.
\end{eqnarray}
Multiplying equation (\ref{Sol7}) by $S= \dot{a}+\dot{b}x$ to clear the denominators, we obtain the following polynomial in $x$ with the constant, can be written as

\begin{eqnarray}
	0&=&-3\dot{a}-2\frac{\dot{k}}{k}a+\left(-3\dot{b}-2\frac{\dot{k}}{k}b\right)x\nonumber\\&&+k^2\left(4a^2\dot{b}-8ab\dot{a}\right)+k^2\left(-8ab\dot{b}+4b^2\dot{a}\right)x,
\end{eqnarray}
and equating the coefficients of the constant term and the coefficient of $x$ separately yields

\begin{eqnarray}
	0&=&-3\dot{a}-2\frac{\dot{k}}{k}a+k^2\left(4a^2\dot{b}-8ab\dot{a}\right)\label{Sol9}\\ \nonumber\\
	0&=&\left(-3\dot{b}-2\frac{\dot{k}}{k}b\right)+k^2\left(-8ab\dot{b}+4b^2\dot{a}\right). \label{Sol10}
\end{eqnarray}

We introduce $p(t)=\frac{1}{k^2}$, then equation (\ref{Sol9}) and (\ref{Sol10}) become first-order linear differential equations in $p$. 

Since $p=k^{-2}$, we obtain $\dot p=-2(\dot k/k^3)$, and yields

\begin{eqnarray}
	0&=&a\dot{p}-3\dot{a}p+4a^2\dot{b}-8ab\dot{a}\\ \nonumber\\
	0&=&b\dot{p}-3\dot{b}p-8ab\dot{b}+4b^2\dot{a}
\end{eqnarray}
by solving the simultaneous equations, we obtain

\begin{equation}
	p(t)=-4a(t)b(t),
\end{equation}
hence $k(t)=\pm\frac{1}{\sqrt{-4a(t)b(t)}}$

Therefore, we have:

\begin{eqnarray}
	L&=& a(t)+b(t)x\\ \nonumber \\
	G&=&\pm \frac{\dot{a(t)}+\dot{b(t)}x}{\sqrt{-4a(t)b(t)}}
\end{eqnarray}

where $-4a(t)b(t)>0$ and $S=\dot{a}+\dot{b}x \neq 0$

\section{Solution via First-Order ODE Approach $p=L_t$\label{ODEP}}

In the equation (\ref{master}), we impose 
\begin{equation}
	2L_{tt}L - 3L_t^2 - 2\frac{G_t}{G}L_tL = 0,\label{S2}
\end{equation}

and let
\begin{equation}
	p = L_t, \quad \text{so} \quad L_{tt} = p \frac{dp}{dL}\label{S2a}
\end{equation}

Substituting equation (\ref{S2a}) into (\ref{S2}), yields 
\begin{equation}
	2L \cdot p \frac{dp}{dL} - 3p^2 - 2\frac{G_t}{G}L p = 0\label{S2b}
\end{equation}

Multiply through equation (\ref{S2b}) by \( G^2 \) and divide by \( p \) (assuming \( p \neq 0 \)):
\begin{equation}
	2L \frac{dp}{dL} - 3p - 2\frac{G_t}{G}L = 0.\label{ODE}
\end{equation}

This is a first-order linear ordinary differential equation for \(p(L)\). Rearranging equation (\ref{ODE}), we obtain
\begin{equation}
	\frac{dp}{dL}-\frac{3}{2L}p=\frac{G_t}{G}.\label{S2c}
\end{equation}

Using the integrating factor in equation (\ref{S2c}), 
\begin{equation}
	\mu(L) = \exp\left(-\int \frac{3}{2L} \, dL\right) = L^{-3/2},
\end{equation}
then we have 
\begin{equation}
	L^{-3/2}p = \frac{G_t}{G} \int L^{-3/2} \, dL + f(x),
\end{equation}

where $f(x)$ is an arbitrary function arising from integration. For simplicity we consider the particular case $f(x)=0$.
This restriction allows the construction of an explicit exact solution while preserving the essential dissipative features of the model. We obtain

\begin{equation}
	L^{-3/2}p = -2\frac{G_t}{G}L^{-1/2},
\end{equation}

hence

\begin{equation}
	L_t = -2\frac{G_t}{G}L \label{L}
\end{equation}

Solving equation (\ref{L}), we obtain

\begin{equation}
	L(x,t)=\frac{C(x)}{G(x,t)^2}.\label{L1}
\end{equation}

Since the master equation (\ref{master}) is a highly nonlinear partial differential equation involving coupled derivatives of $ L$ and $ G$, obtaining exact analytical solutions in its general form is extremely difficult. We therefore introduce the auxiliary choice

\begin{equation}
	G=\frac{1}{a(t)},
	\label{G}
\end{equation}
where $a(t)$ is arbitrary. This assumption removes the spatial derivatives of G, significantly simplifying the nonlinear coupling while retaining sufficient freedom to construct physically meaningful radiating stellar models. The reduced equation admits exact closed-form solutions for $L(x,t)$, enabling a detailed investigation of the corresponding gravitational and physical properties.

Substituting equation (\ref{L1}) and (\ref{G}) into (\ref{master}), we obtain

\begin{equation}
	4a^4 C'(x)\big(2\sqrt{x}C(x)a_t(t) + 3xC'(x) - 2C(x)\big) = 0.\label{S2d}
\end{equation}

\section*{Case (i): The Constraint, $C'(x) = 0$}

In equation (\ref{S2d}), if  $C'(x) = 0$, then $C(x) = C_0$ constant, and
\begin{equation}
	L(x,t) = C_0 a(t)^2,\quad a(t)\ \text{arbitrary}.
\end{equation}
\section*{Case (ii): The Constraint,	 $ C'(x) \neq 0$}
if  $ C'(x) \neq 0$:
\begin{equation}
	2\sqrt{x}C \dot{a} + 3xC' - 2C = 0.
\end{equation}
To obtain an explicit closed-form solution, we consider the particular case in which the temporal derivative is constant,  $\dot{a} = k$. where $k$ is constant. $ a(t) = kt + a_1$.
Then:
\begin{equation}
	\frac{C'}{C} = \frac{2}{3}\left(\frac{1}{x} - \frac{k}{\sqrt{x}}\right).\label{S2e}
\end{equation}
Integrating equation (\ref{S2e}), we obtain
\begin{equation}
	\ln C = \frac{2}{3}\ln x - \frac{4k}{3}\sqrt{x} + \ln C_0,
\end{equation}
therefore
\begin{equation}
	C(x) = C_0 x^{2/3} e^{-\frac{4k}{3}\sqrt{x}}.
\end{equation}

The above equations yield 
\begin{equation}
	L(x,t) = C_0 x^{2/3} e^{-\frac{4k}{3}\sqrt{x}}(kt + a_1)^2.\label{SL}
\end{equation}
Recall that $x=r^2$.
Since $L(r,t)\propto r^4/3$
, it follows that $ B(r,t)=L^{-1}(r,t)\propto r^{-4/3}$.
Hence, the metric potential $ B(r,t)$ diverges as $ r\rightarrow 0$, and the exact solution is singular at the centre. Consequently, the solution cannot represent a complete stellar interior extending from the centre to the surface and is instead interpreted as describing a stellar envelope.

\section{Complexity \label{complexit}}

Following the definition introduced by Bogadi \textit{et al.}~\cite{BogadiGovenderMoyo2022a}, the complexity factor for a dissipative anisotropic fluid distribution is given by

\begin{equation}
	Y_{TF}
	=
	8\pi\Pi
	+
	\frac{4\pi}{(rB)^3}
	\int_{0}^{r}
	(rB)^3
	\left(
	\mu'
	-
	3qB\frac{\dot{B}}{A}
	\right)
	dr,
	\label{complexity}
\end{equation}

where

\[
\Pi=P_r-P_{\perp}
\]

denotes the local pressure anisotropy.

For convenience, we decompose the complexity factor into three physically distinct contributions,

\begin{align}
	Z_{1} &= 8\pi\left(P_r-P_{\perp}\right), \label{Z1}\\[2mm]
	Z_{2} &= \frac{4\pi}{(rB)^3}
	\int_{0}^{r}
	(rB)^3
	\mu'\,dr,
	\label{Z2}\\[2mm]
	Z_{3} &= -\frac{12\pi}{(rB)^3}
	\int_{0}^{r}
	(rB)^3
	qB\frac{\dot{B}}{A}\,dr.
	\label{Z3}
\end{align}

Hence, the complexity factor can be written in the compact form

\begin{equation}
	Y_{TF}=Z_{1}+Z_{2}+Z_{3}.
\end{equation}

The three contributions have clear physical interpretations:
\begin{itemize}
	\item [$\bullet$] $Z_{1}$ represents the contribution from the local anisotropy of the fluid pressure.
	\item [$\bullet$] $Z_{2}$ measures the contribution arising from the inhomogeneity of the energy density.
	\item [$\bullet$] $Z_{3}$ accounts for the dissipative effects associated with the radial heat flux.
\end{itemize}

Therefore, the complexity of the self-gravitating system is determined by the combined influence of pressure anisotropy, density inhomogeneity, and dissipative processes.

\section{Analysis of the Physical Properties of the Radiating Star} \label{Analysis} 
Using the transformations introduced in Section~\ref{Stellar},
\begin{equation}
	x=r^{2}, \qquad L=B^{-1}, \qquad G=LA,
\end{equation} 
together with the auxiliary assumption 
\begin{equation}
	G=\frac{1}{a(t)}, 
\end{equation} introduced in Section~\ref{ODEP}, the exact solution obtained in Section~\ref{ODEP} can be written in a form suitable for physical analysis.

The exact solution for the metric function $L$ obtained in
Section~\ref{ODEP} is given by
\begin{equation}
	L(x,t)=C_{0}\,x^{2/3}\exp\!\left(-\frac{4k}{3}\sqrt{x}\right)(kt+a_{1})^{2}.
\end{equation}

Since \(x=r^{2}\), it follows that \(\sqrt{x}=r\), and hence the above solution can be expressed in terms of the radial coordinate as
\begin{equation}
	L(r,t)=C_{0}\,r^{4/3}\exp\!\left(-\frac{4kr}{3}\right)(kt+a_{1})^{2}.
\end{equation}

Using the relation \(L=B^{-1}\), we obtain the gravitational potential
\begin{equation}
	B(r,t)=\frac{\exp\!\left(\frac{4kr}{3}\right)}
	{C_{0}\,r^{4/3}(kt+a_{1})^{2}}.
\end{equation}

Furthermore, since \(G=LA=1/a(t)\), the second gravitational potential is given by
\begin{equation}
	A(r,t)=\frac{B(r,t)}{a(t)}
	=\frac{\exp\!\left(\frac{4kr}{3}\right)}
	{a(t)\,C_{0}\,r^{4/3}(kt+a_{1})^{2}}.
\end{equation}
Since the metric function $B(r,t)$ contains the factor $r^{-4/3}$
, the present exact solution is interpreted as describing the envelope region of the radiating star rather than the complete interior extending to the centre $r=0$. Accordingly, the physical analysis below is performed over the regular envelope region, where all physical variables are finite and well behaved.

The metric functions $A(r,t)$ and $B(r,t)$ completely determine
the interior geometry of the radiating stellar configuration. In
the following subsection, we investigate the behaviour of the
corresponding matter variables and kinematical quantities to
assess the physical behaviour of the obtained solution.

For convenience, define $T(t) = kt + a_1$ and $C(r) = C_0 e^{-\frac{4k}{3}r}$, 
then the field equations (\ref{14a})--(\ref{14d}) become

\begin{eqnarray}
	\mu &= & 12k^2 a(t)^2 C^2 T^2 r^{8/3}
	\\ \nonumber&&	+ \frac{16}{27} C^3 T^6(-2 - 3kr + 9k^2r^2 - 4k^3r^3)\\
	\nonumber&&+ \frac{8}{3} C^2 T^4(r^{2/3} + k r^{5/3} - 2k^2 r^{8/3}),\\ \nonumber\\
	P_r &= &\big[-8k^2 a(t)^2 C^2 T^2 r^{8/3} + 4k a(t)\dot a(t) C^2 T^3 r^{8/3}\big]\\
	\nonumber&& - \frac{16k}{3} C^2 T^4(r^{5/3} - k r^{8/3}),\\ \nonumber\\
	P_\perp &= &\big[-8k^2 a(t)^2 C^2 T^2 r^{8/3} + 4k a(t)\dot a(t) C^2 T^3 r^{8/3}\big]\\
	\nonumber&& + \frac{8}{9} C^2 T^4(2r^{2/3} - k r^{5/3} + 2k^2 r^{8/3}),\\ \nonumber\\
	q &= &-\frac{16k}{3} a(t) C^3 T^5(r^3 - k r^4).
\end{eqnarray}

To investigate the physical behaviour of the obtained exact solution, we analyse the evolution of the matter variables and the complexity factor. The energy density, radial pressure, tangential pressure, and heat flux are examined as functions of both the radial coordinate and time. In addition, the complexity factor is decomposed into its anisotropic, density-inhomogeneity, and dissipative contributions to identify the dominant physical mechanisms governing the evolution of the radiating envelope. Unless otherwise stated, the numerical analysis is performed using the parameter values $k=-3.9$ and $a_1=10000$. These parameter values yield well-behaved matter variables throughout the prescribed physical envelope domain and are used consistently in the numerical analysis.

\subsection{Physical Domain of the Exact Solution} \label{physicaldomain} The exact solution obtained in the preceding section is not regular at the centre \(r=0\). In particular, since

\begin{equation} L(x,t)=C_0x^{2/3} \exp\!\left(-\frac{4k}{3}\sqrt{x}\right)(kt+a_1)^2, \end{equation} it follows that

\begin{equation} B(r,t)=\frac{1}{L} = \frac{\exp(4kr/3)} {C_0r^{4/3}(kt+a_1)^2}, 
\end{equation}

which diverges as \(r\rightarrow 0\). Consequently, the solution cannot represent a complete stellar interior extending from the centre to the surface. Instead, it is interpreted as describing the envelope region of a radiating star. The physical analysis is therefore restricted to the finite stellar-envelope domain 
\begin{equation} 
	r_0\leq r\leq r_\Sigma, \qquad r_0>0, \label{x}
\end{equation} where \(r_0\) denotes the inner boundary of the envelope and \(r_\Sigma\) denotes the stellar surface. In terms of the transformed coordinate \(x=r^2\), the corresponding domain is \[ x_0\leq x\leq x_\Sigma, \qquad x_0=r_0^2, \qquad x_\Sigma=r_\Sigma^2. \] For the numerical analysis presented in this paper, all plots are generated over the finite envelope region defined by Eq.~(\ref{x}). The interval is chosen so that the metric functions and thermodynamic variables remain finite, continuous, and physically well behaved throughout the domain. No physical interpretation is assigned to the region \[ 0\leq r<r_0, \] since the exact solution is not valid there. Throughout the analysis, we adopt \[ r_0=0.1~{\rm km}, \qquad r_\Sigma=1.0~{\rm km}. \] The singular behaviour of the metric potential \(B(r,t)\propto r^{-4/3}\) at the centre further reinforces the interpretation of the solution as a stellar-envelope model. Similar core--envelope descriptions have been employed successfully in relativistic stellar modelling, where the envelope is matched smoothly to a regular core at an internal boundary \(r=r_0>0\) and to the exterior spacetime at the stellar surface \(r=r_\Sigma\) \cite{TakisaMaharajMulangu2019,KhuntThomasVinodkumar2023}. Accordingly, all physical quantities and complexity variables presented in this work are evaluated exclusively within the admissible envelope region \(r_0\leq r\leq r_\Sigma\). This ensures that the analysis is confined to the physically meaningful domain for which the exact solution is valid.

\subsection{Gravitational Complexity \label{complexit2}}
As discussed in Sec.~\ref{physicaldomain}, the exact solution obtained in this work is singular at the centre, $r=0$, and is therefore interpreted as describing the radiating envelope of a core--envelope configuration. The physical domain of the exact solution is consequently restricted to
\begin{equation}
	r_0\leq r\leq r_{\Sigma},
	\qquad r_0>0,
	\label{physicaldomaincomplexity}
\end{equation}
where $r_0$ denotes the inner boundary of the envelope and $r_{\Sigma}$ represents the stellar surface.

For $r\geq r_0$, the integral appearing in Eq.~\eqref{complexity} can be separated into a contribution from the core and a contribution from the exact envelope solution according to
\begin{equation}
	\int_{0}^{r}(\cdots)\,dr'
	=
	\int_{0}^{r_0}(\cdots)\,dr'
	+
	\int_{r_0}^{r}(\cdots)\,dr'.
\end{equation}
The first integral represents the contribution inherited from the core, whereas the second is evaluated entirely using the exact solution derived in this paper. Defining the core contribution by
\begin{equation}
	\mathcal{C}_0(t)
	=
	4\pi
	\int_{0}^{r_0}
	(r'B)^3
	\left(
	\mu'
	-
	3qB\frac{\dot{B}}{A}
	\right)
	dr',
	\label{corecontribution}
\end{equation}
the complexity factor in the envelope can be written as
\begin{eqnarray}
	Y_{TF}(r,t)
	&=&
	8\pi\Pi(r,t)
	+
	\frac{\mathcal{C}_0(t)}
	{[rB(r,t)]^3}
	\nonumber\\&&+
	\frac{4\pi}{[rB(r,t)]^3}
	\int_{r_0}^{r}
	[r'B(r',t)]^3
	\biggl[
	\mu'(r',t)
	\nonumber\\&&	-
	3q(r',t)
	B(r',t)
	\frac{\dot{B}(r',t)}
	{A(r',t)}
	\biggr]
	dr'.
	\label{envelopecomplexity}
\end{eqnarray}

For convenience, the complexity factor is decomposed into three physically distinct contributions,
\begin{eqnarray}
	Z_1(r,t)
	&=&
	8\pi\left[P_r(r,t)-P_{\perp}(r,t)\right],
	\label{Z1}\\[2mm]
	Z_2(r,t)
	&=&
	\frac{\mathcal{C}_{2,0}(t)}
	{[rB(r,t)]^3}
	\\ \nonumber&&	+
	\frac{4\pi}{[rB(r,t)]^3}
	\int_{r_0}^{r}
	[r'B(r',t)]^3
	\mu'(r',t)\,dr',
	\label{Z2}\\[2mm]
	Z_3(r,t)
	&=&
	\frac{\mathcal{C}_{3,0}(t)}
	{[rB(r,t)]^3}
	-
	\frac{12\pi}{[rB(r,t)]^3}
	\int_{r_0}^{r}
	\biggl[	[r'B(r',t)]^3
	\\ \nonumber&&	q(r',t)
	B(r',t)
	\frac{\dot{B}(r',t)}
	{A(r',t)}
	dr'\biggr],
	\label{Z3}
\end{eqnarray}
where $\mathcal{C}_{2,0}(t)$ and $\mathcal{C}_{3,0}(t)$ denote the contributions inherited from the core to the density-inhomogeneity and dissipative parts of the complexity factor, respectively.

With these definitions, the total complexity factor retains the compact decomposition
\begin{equation}
	Y_{TF}=Z_1+Z_2+Z_3.
	\label{complexitydecomposition}
\end{equation}

The three contributions have distinct physical interpretations. The term $Z_1$ represents the contribution of local pressure anisotropy, and therefore measures the difference between the radial and tangential pressures. The term $Z_2$ describes the contribution associated with the radial inhomogeneity of the energy density. The term $Z_3$ represents the contribution arising from dissipative heat transport through the radial heat flux.

Thus, within the exact envelope domain, the gravitational complexity is determined by the combined effects of pressure anisotropy, density inhomogeneity, and dissipative heat transport. The separation introduced in Eqs.~\eqref{Z1}--\eqref{Z3} allows the relative importance of these mechanisms to be investigated throughout the evolution of the radiating envelope. For the physical analysis of $Z_2$ and $Z_3$, we adopt the simplifying modelling assumption
\begin{equation}
	\mathcal{C}_{2,0}(t)=\mathcal{C}_{3,0}(t)=0.
\end{equation}

This assumption is introduced solely to isolate the contribution of the exact envelope solution derived in the present work. Since the exact solution describes only the stellar envelope and not the central core, the full complexity factor would generally contain additional contributions inherited from the interior core configuration through the constants $	\mathcal{C}_{2,0}(t)$ and $	\mathcal{C}_{3,0}(t)$.

The purpose of setting these quantities equal to zero is therefore not to imply the absence of density inhomogeneity or dissipative effects within the core. Rather, it enables the complexity generated by the exact envelope solution itself to be identified and studied independently of any particular choice of core model. In this way, the influence of pressure anisotropy, density gradients, and dissipative heat transport arising uniquely from the envelope can be examined without introducing additional free parameters associated with a specific core geometry.

Consequently, the results presented in this paper should be interpreted as describing the intrinsic complexity structure of the radiating envelope. In a complete core-envelope model, nonvanishing functions  $	\mathcal{C}_{2,0}(t)$ and $	\mathcal{C}_{3,0}(t)$ would contribute additive corrections to $Z_2$ and $Z_3$, respectively.

The qualitative behaviour identified here, particularly the increasing importance of the dissipative contribution near the stellar surface, is expected to remain informative, although the precise magnitudes of the complexity components would depend on the chosen core solution. The investigation of matched core-envelope configurations with nonzero core contributions remains an interesting topic for future work. Under this assumption, the density-inhomogeneity and dissipative contributions are evaluated
solely from the exact envelope solution over the physical domain
$r_0\leq r\leq r_\Sigma$.

\subsection{The expansion scalar}

For the exact solution obtained in Section~\ref{ODEP}, the expansion scalar given by Eq.~(\ref{Theta1}) takes the form
\begin{equation}
	\Theta
	=
	-6k\,a(t)\,C_{0}\,r^{4/3}
	\exp\!\left(-\frac{4kr}{3}\right)
	(kt+a_{1}).\label{Theta2}
\end{equation}
For the parameter values adopted in the numerical analysis, \(k=-3.9\), \(a_1=10000\), and assuming \(a(t)>0\), \(C_0>0\), \(r>0\), and \(kt+a_1>0\), all factors appearing on the right-hand side of Eq.(\ref{Theta2}) are positive. Since \(k=-3.9<0\), it follows that \[ -6k>0. \] Consequently, \[ \Theta>0 \] throughout the physical envelope domain and over the temporal interval considered.
A positive expansion scalar indicates that the proper volume of neighbouring fluid elements increases with time. Thus, for the chosen parameter values, the exact solution describes an \emph{expanding radiating stellar envelope}. Moreover, the non-zero heat flux indicates that this expansion takes place in the presence of dissipative energy transport, with energy being carried outward from the stellar matter towards the exterior Vaidya spacetime. Thus, the positive expansion scalar provides a direct kinematical
indication that the proper volume of a comoving fluid element
increases with time. The model therefore describes an expanding,
rather than contracting, radiating envelope for the parameter values
considered.

\subsection{Mass Function}

The boundary mass function given by Eq.~(\ref{massLG}) was evaluated using the exact solution obtained in Section~\ref{ODEP}.

\begin{equation}
	m(v) = \left[ \frac{4\,e^{\frac{4kr}{3}}} {9C_0\,r^{1/3}(kt+a_1)^2} \left( 2+2kr+5k^2r^2 \right) \right]_{\Sigma}. 
\end{equation}

For the parameter values adopted in the numerical analysis, the mass remains positive throughout the physical envelope region. This indicates that the obtained radiating configuration possesses a positive gravitational mass throughout the prescribed physical domainduring its evolution.

\begin{figure}[htbp]
	\centering
	\includegraphics[scale=0.5]{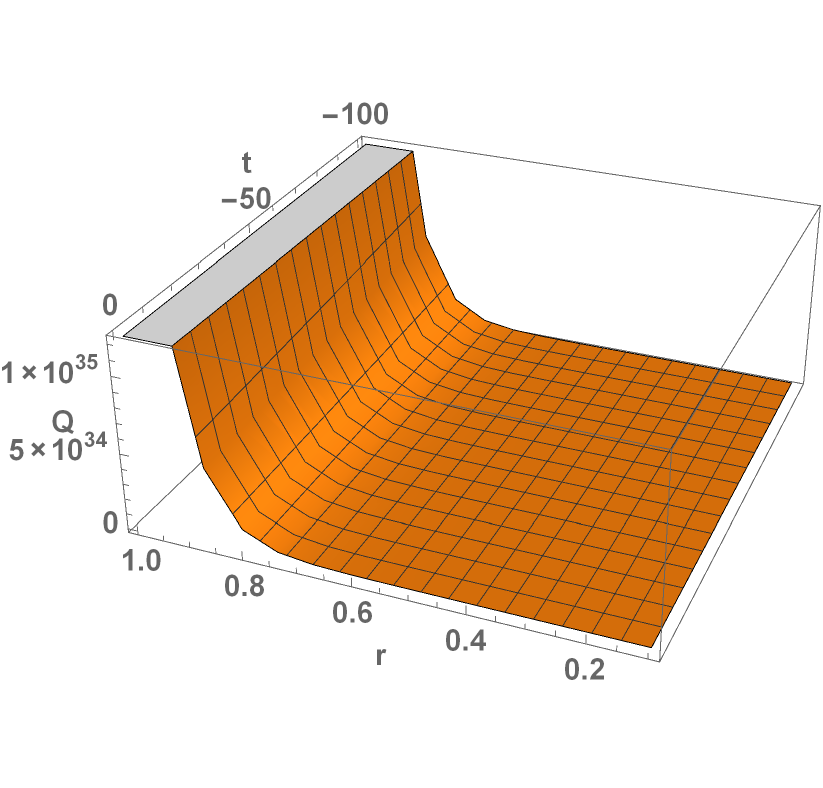}\caption{ $q(r,t)$ as a function of the radial coordinate $r$
		and temporal coordinate $t$.}, \label{fig1}
\end{figure}

Figure \ref{fig1} illustrates the evolution of the heat flux $q(r,t)$. For the parameter values considered in this study, the heat flux remains positive throughout the stellar envelope. This indicates that thermal energy is transported outward from the stellar interior toward the exterior Vaidya spacetime, consistent with the interpretation of the configuration as a radiating star. The positive heat flux therefore provides evidence of dissipative energy transport from the stellar matter to the exterior. Moreover, the magnitude of the heat flux varies with both the radial coordinate and time, indicating that the dissipative effects evolve dynamically during the evolution of the stellar configuration. In particular, the increase in the magnitude of $q(r,t)$ over the considered temporal interval indicates that heat transport becomes increasingly significant as the configuration evolves. The heat flux also varies smoothly with the radial coordinate over the region considered, with no discontinuities or abrupt changes, indicating that the dissipative contribution remains well behaved throughout the stellar envelope.

\begin{figure}[htbp]
	\centering
	\includegraphics[scale=0.5]{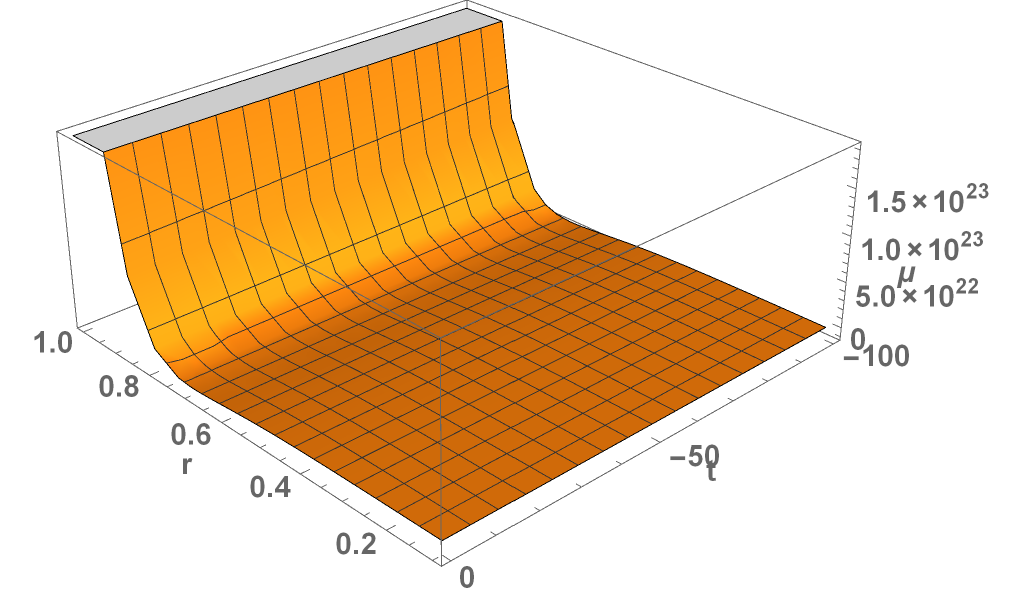}\caption{Energy density $\mu(r,t)$ as a function of the radial coordinate $r$ and temporal coordinate $t$.}, \label{fig2}
\end{figure}

Figure \ref{fig2} displays the energy density $\mu(r,t)$. The density remains positive throughout the stellar envelope, satisfying one of the basic physical requirements for a realistic stellar model. The density decreases monotonically from the inner boundary of the envelope towards the surface, indicating that matter remains more concentrated in the deeper layers of the stellar configuration. As time evolves, the density gradually changes, reflecting the redistribution of matter caused by dissipative radiative processes and the dynamical evolution of the envelope.

\begin{figure}[htbp]
	\centering
	\includegraphics[scale=0.5]{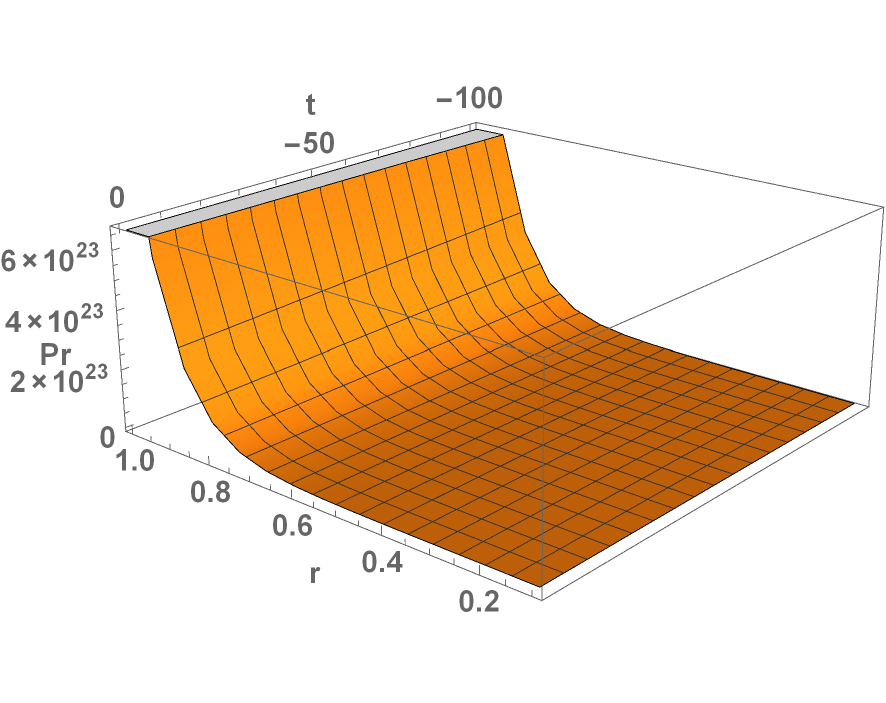}\caption{Radial pressure $P_r(r,t)$ as a function of the radial coordinate $r$ and temporal coordinate $t$.},\label{fig3}
\end{figure}

Figure~\ref{fig3} illustrates the radial pressure $P_r(r,t)$. The
radial pressure remains finite throughout the physical stellar
envelope domain and decreases with increasing radius, indicating a
well-behaved radial pressure profile. As the stellar envelope
evolves, the pressure changes continuously without exhibiting
oscillatory behaviour or discontinuities.

\begin{figure}[htbp]
	\centering
	\includegraphics[scale=0.5]{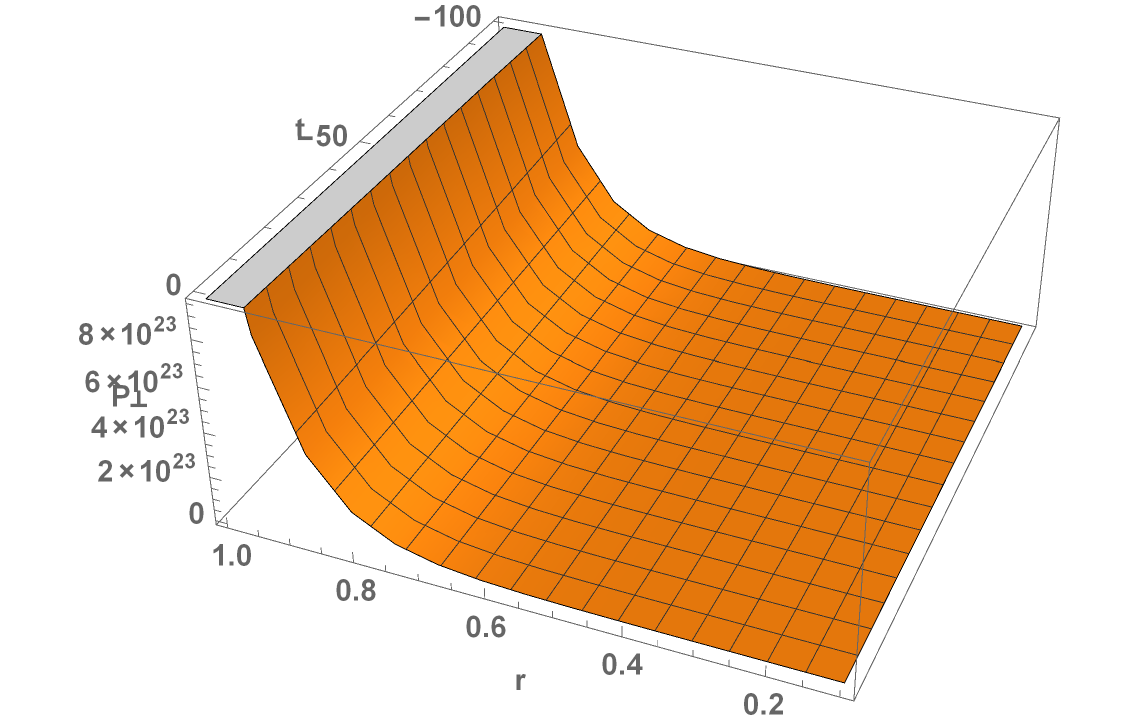}\caption{Tangential pressure $P_{\perp}(r,t)$ as a function of the radial coordinate $r$ and temporal coordinate $t$.}, \label{fig4}
\end{figure}

Figure \ref{fig4} illustrates the tangential pressure $P_\perp(r,t)$. The tangential pressure exhibits behaviour similar to the radial pressure but differs quantitatively, demonstrating the presence of local pressure anisotropy. The difference between the two pressures contributes directly to the gravitational complexity of the system through the anisotropic component $Z_{1}$.

The smooth variation of $P_\perp$ indicates that the anisotropic stresses remain well behaved throughout the evolution of the radiating envelope.

\begin{figure}[htbp]
	\centering
	\includegraphics[scale=0.5]{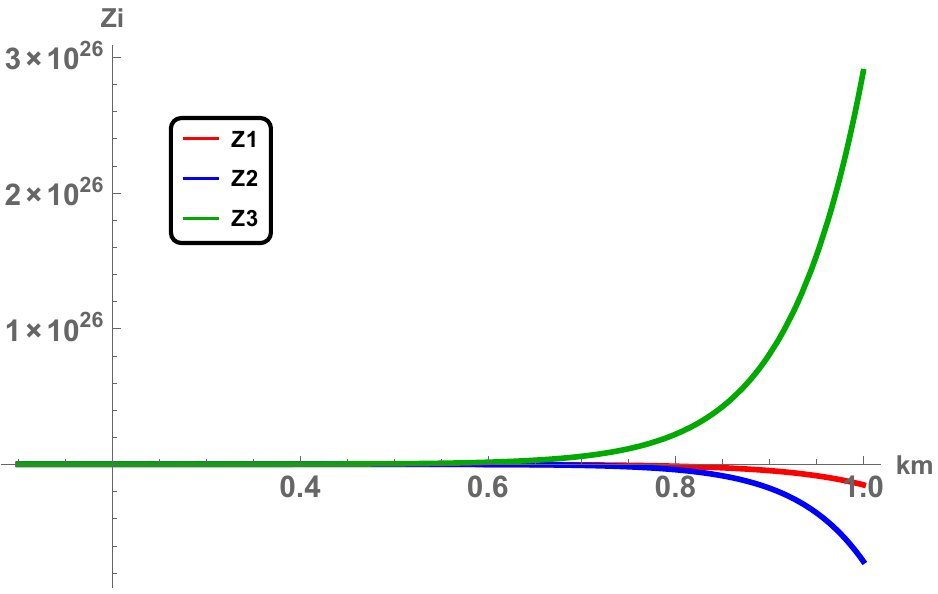}\caption{Behaviour of the complexity components $Z_1$, $Z_2$, and $Z_3$ as functions of the radial coordinate $r$ for fixed temporal value  $t=-100$}, \label{fig5}
\end{figure}

Figure \ref{fig5} illustrates the radial behaviour of the three contributions to the complexity factor, namely the anisotropic component $Z_{1}$, the density inhomogeneity component $Z_{2}$, and the dissipative component $Z_{3}$, at the fixed time  $ t=-100$. Throughout most of the stellar envelope, the complexity is primarily governed by the combined effects of pressure anisotropy and density inhomogeneity. However, as the surface of the star is approached $(r \gtrsim 0.8 km)$, the dissipative contribution  $Z_{3}$
increases rapidly and becomes the dominant positive component. In contrast, both the anisotropic contribution  $Z_{1}$
and the density-gradient contribution  $Z_{2}$ remain negative in this region, with  $Z_{2}$ providing the larger negative contribution. This behaviour indicates that heat dissipation becomes the principal mechanism controlling the gravitational complexity near the stellar boundary, while the effects of anisotropy and density inhomogeneity partially counteract its influence. The competition among these three contributions demonstrates that the complexity of the evolving radiating configuration is determined by their combined rather than individual effects.

\begin{figure}[htbp]
	\centering
	\includegraphics[scale=0.5]{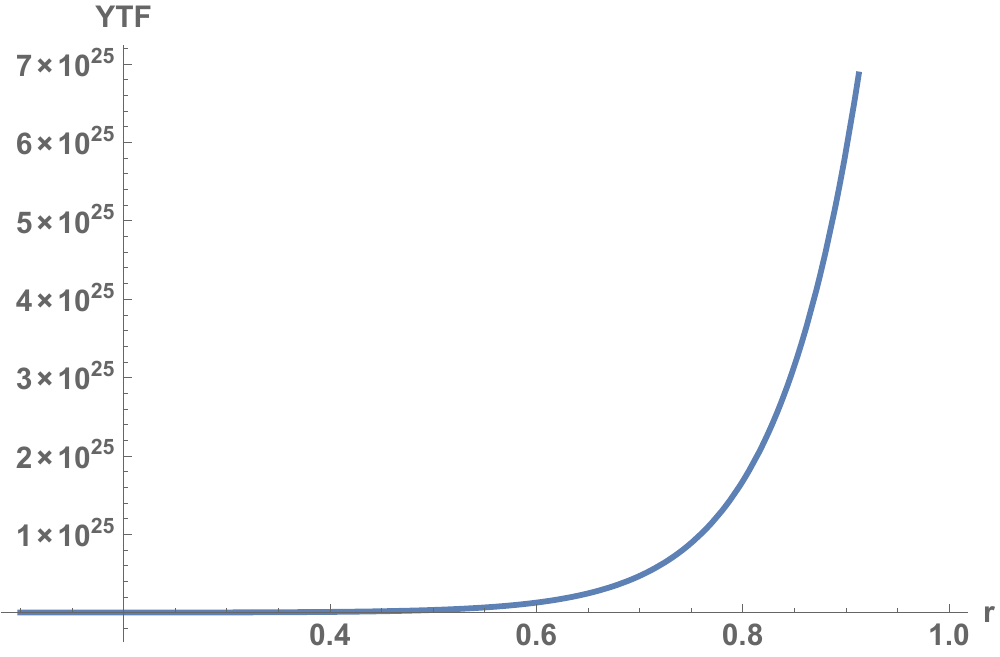}\caption{Evolution of the complexity factor $Y_{TF}$ with respect to the radial coordinate $r$ for fixed temporal value  $t=-100$.}, \label{fig6}
\end{figure}

Figure \ref{fig6} presents the total complexity factor $Y_{TF}$ as a function of the radial coordinate at the same instant, $t=-100$. The complexity factor remains finite and well behaved throughout the prescribed physical envelope domain, demonstrating that the complexity is regular away from the central singularity. A noticeable increase occurs for $(r \gtrsim 0.8,\mathrm{km})$, where the positive dissipative contribution shown in Figure \ref{fig5} outweighs the negative contributions arising from pressure anisotropy and density inhomogeneity. Consequently, the total complexity increases sharply near the stellar surface, indicating that dissipative heat transport plays a dominant role in determining the gravitational complexity in the outer region of the radiating envelope.

\begin{figure}[htbp]
	\centering
	\includegraphics[scale=0.5]{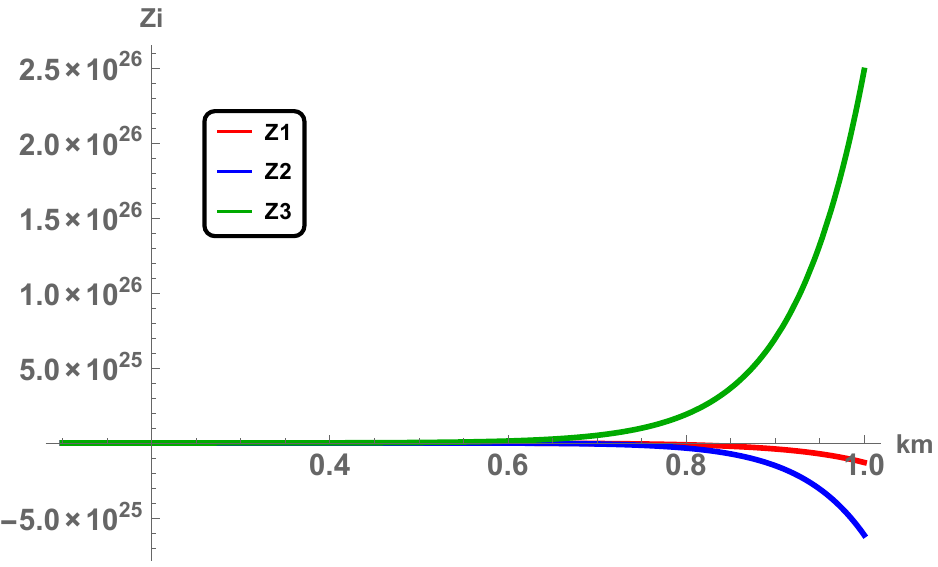}\caption{Behaviour of the complexity components $Z_1$, $Z_2$, and $Z_3$ as functions of the radial coordinate $r$ for fixed temporal value  $t=0$}, \label{fig7}
\end{figure}

\begin{figure}[htbp]
	\centering
	\includegraphics[scale=0.5]{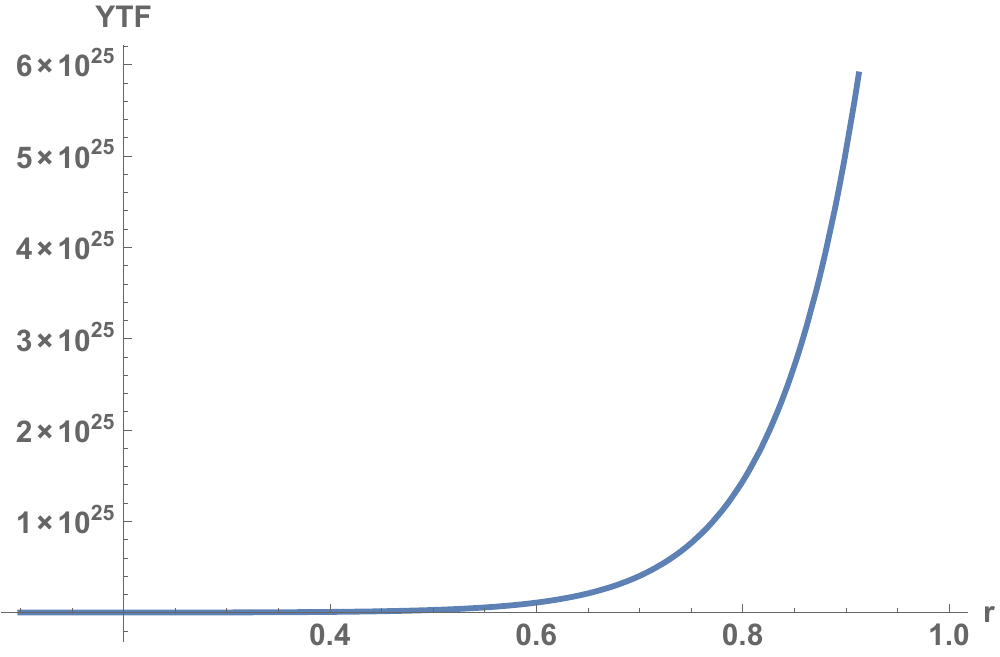}\caption{Evolution of the complexity factor $Y_{TF}$ with respect to the radial coordinate $r$ for fixed temporal value  $t=0$.}, \label{fig8}
\end{figure}

Figures \ref{fig7} and \ref{fig8} present the same quantities at
$t=0$, allowing a direct comparison with the corresponding profiles
at $t=-100$ shown in Figures \ref{fig5} and \ref{fig6}. The relative
magnitudes of the three complexity components change with time,
demonstrating that gravitational complexity evolves dynamically
during the expansion of the radiating stellar envelope. This
time-dependent behaviour highlights the changing contributions of
pressure anisotropy, density inhomogeneity, and dissipative heat
transport to the overall gravitational complexity.

\begin{figure}[htbp]
	\centering
	\includegraphics[scale=0.5]{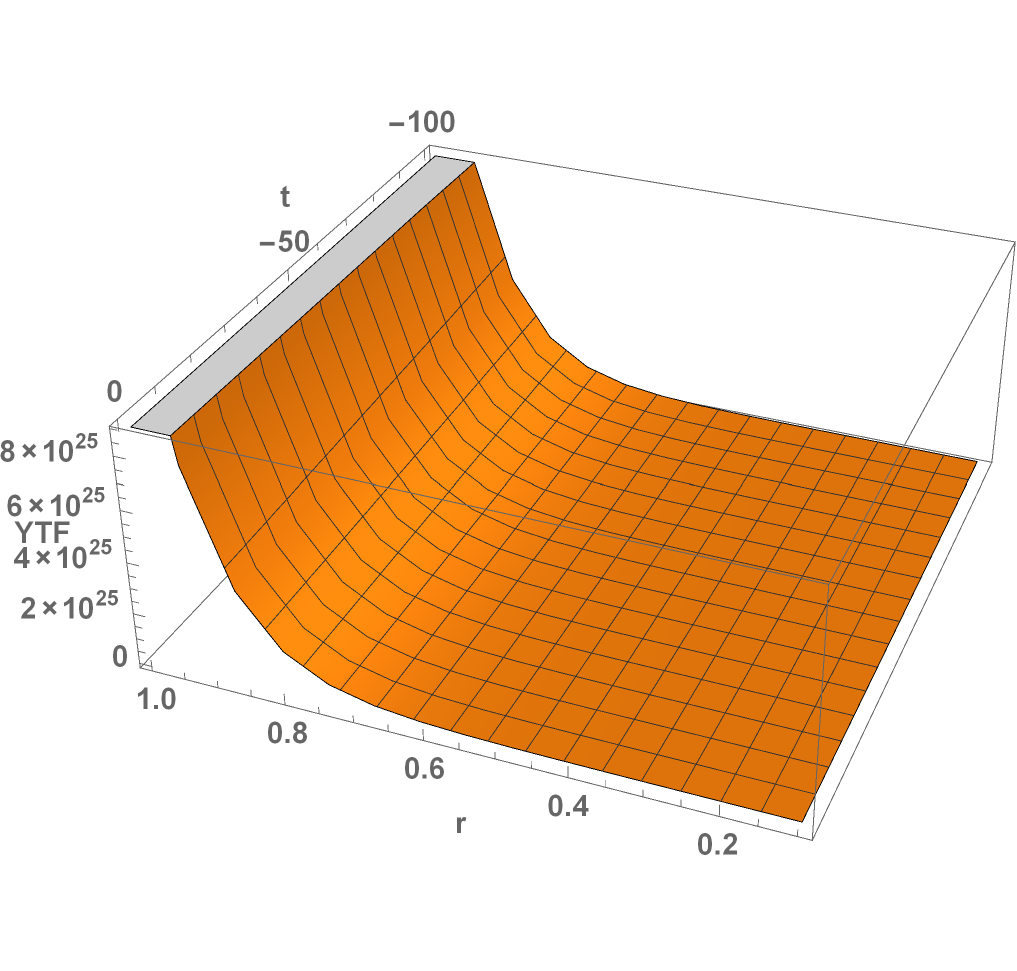}\caption{Evolution of the complexity factor $Y_{TF}$ with respect to the radial coordinate $r$ and temporal $t$ coordinates.}, \label{fig9}
\end{figure}

Figure \ref{fig9} illustrates the full spacetime evolution of the total complexity factor $Y_{TF}(r,t)$. The three-dimensional profile demonstrates that the complexity evolves smoothly with both radius and time. This indicates that the gravitational complexity evolves continuously during the evolution of the radiating envelope without exhibiting any pathological behaviour, further supporting the physical consistency of the obtained exact solution.

\begin{figure}[htbp]
	\centering
	\includegraphics[scale=0.5]{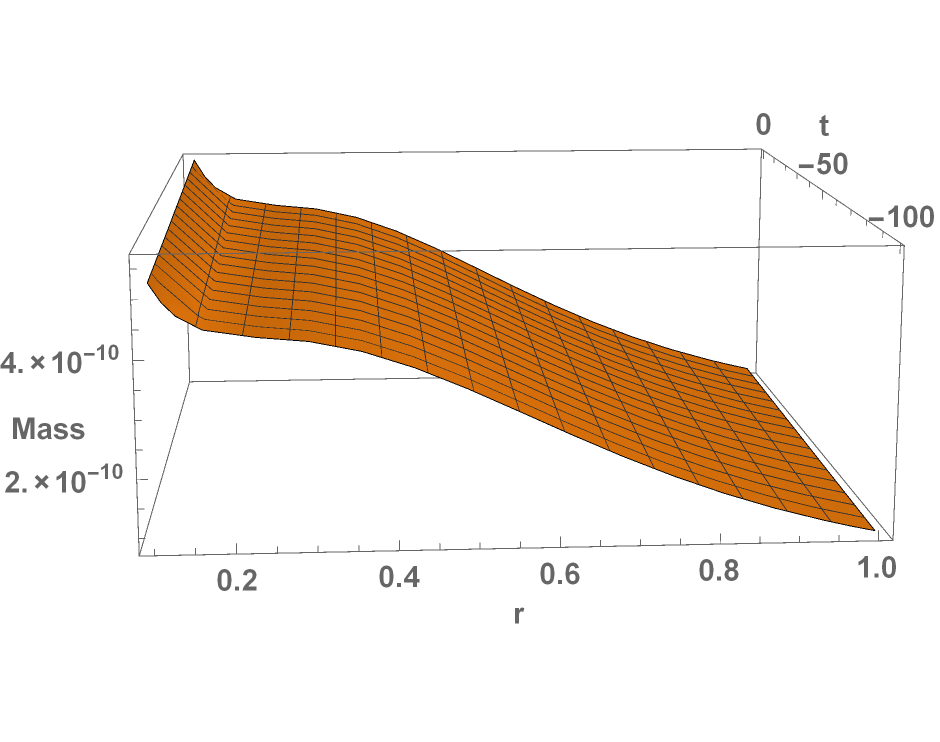}\caption{ $Mass$ function with respect to the radial coordinate $r$ and temporal $t$ coordinates.}, \label{fig10}
\end{figure}

Figure~\ref{fig10} shows that the mass remains finite and positive throughout the evolution, while its smooth dependence on both radius and time is consistent with the interpretation of the model as well-behaved radiating stellar envelope.

\section{Conclusion}\label{Conclusion}

In this paper, we have investigated the gravitational behaviour of
a shear-free radiating stellar envelope undergoing dissipative
evolution. By introducing suitable transformations, the nonlinear
junction condition was reduced to a more tractable form, enabling
the construction of two classes of exact solutions. An explicit
radiating envelope model was subsequently obtained in closed form
and analysed in detail.

Unlike previous studies in which the physical analysis was mainly
restricted to the radial coordinate, the present work investigates
the evolution of the radiating envelope in both space and time.
The matter variables and the gravitational complexity factor were
therefore analysed as functions of the radial coordinate $r$ and
temporal coordinate $t$. This provides a more comprehensive
description of the dynamical evolution of the configuration and
allows the influence of dissipative processes to be examined
throughout the evolution.

For the parameter values considered, the exact solution exhibits
well-behaved matter variables throughout the physical envelope
domain. The energy density and pressures remain finite, while the
positive heat flux corresponds to outward-directed energy transport
towards the exterior Vaidya spacetime. The expansion scalar is also
positive, $\Theta>0$, showing that the selected exact solution
describes an expanding radiating stellar envelope.

The decomposition of the complexity factor,
\[
Y_{TF}=Z_1+Z_2+Z_3,
\]
shows that the gravitational complexity is governed by the combined
effects of pressure anisotropy, density inhomogeneity, and
dissipative heat transport. In particular, for the parameter values
considered, the dissipative contribution $Z_3$ becomes increasingly
significant towards the outer boundary of the envelope and can
dominate the other contributions. This demonstrates that heat
transport can play an important role in determining the gravitational
complexity of an evolving radiating configuration.

The exact solutions obtained in this work provide additional analytical models for dissipative evolution in shear-free radiating stellar envelopes.
They may also provide useful benchmarks for future investigations of
relativistic stellar evolution and gravitational complexity.

The exact solution obtained is singular at the centre and therefore describes a radiating stellar envelope rather than a complete stellar interior. All physical quantities were analysed within the admissible envelope region $r_0\leq r\leq r_\Sigma$.

As a direction for future research, the transformations introduced
in this paper will be employed to investigate radiating stellar
models satisfying a linear equation of state of the form
\[
p=\alpha\rho-\beta,
\]
with the aim of constructing further exact solutions and analysing
their physical and gravitational properties. In particular, a
systematic investigation of the standard energy conditions will be
undertaken throughout the admissible stellar-envelope domain
\(r_0\leq r\leq r_{\Sigma}\). For the anisotropic dissipative matter
distribution considered here, this analysis will include the weak,
dominant, and strong energy conditions, involving the quantities
\[
\mu,\qquad
\mu+P_r,\qquad
\mu+P_{\perp},\qquad
\mu+P_r+2P_{\perp},
\]
together with the dominant-energy-condition requirements
\[
\mu\geq |P_r|,
\qquad
\mu\geq |P_{\perp}|.
\]
This analysis will provide a more complete assessment of the
physical behaviour of the exact radiating envelope and will help
establish the parameter ranges for which the resulting relativistic
fluid distribution satisfies the standard energy conditions.

\clearpage
\begin{thebibliography}{99}
	\bibitem{OppenheimerSnyder1939} Oppenheimer, J.R., Snyder, H.: Phys. Rev. \textbf{56}, 455 (1939).
	
	\bibitem{Vaidya1951} Vaidya, P.C.: Proc. Indian Acad. Sci. A \textbf{33}, 264 (1951).
	
	\bibitem{Santos1985} Santos, N.O.: Mon. Not. R. Astron. Soc. \textbf{216}, 403 (1985). 
	
	\bibitem{MisnerSharp1964} Misner, C.W., Sharp, D.: Phys. Rev. \textbf{136B}, 571 (1964).
	
	\bibitem{Glass1989} Glass, E.N.: Gen. Relativ. Gravit. \textbf{21}, 733--745 (1989).
	
	\bibitem{ThirukkaneshMoopanarGovender2012} Thirukkanesh, S., Moopanar, S., Govender, M.: Pramana J. Phys. \textbf{79}, 223--232 (2012). 
	
	\bibitem{Ivanov2016} Ivanov, B.V.: Astrophys. Space Sci. \textbf{361}, 18 (2016).
	
	\bibitem{GovenderThirukkanesh2015} Govender, M., Thirukkanesh, S.: Astrophys. Space Sci. \textbf{358}, 39 (2015). 
	
	\bibitem{GovenderReddyMaharaj2009} Govender, M., Reddy, K.P., Maharaj, S.D.: Int. J. Mod. Phys. D \textbf{18}, 1399--1414 (2009).
	
	\bibitem{SifisoandSDMaharaj} Ngubelanga, S.A., Maharaj, S.D.: Adv. Math. Phys. \textbf{2015}, 905168 (2015).
	
	\bibitem{SifisoandSDMaharaj2} Ngubelanga, S.A., Maharaj, S.D., Ray, S.: Astrophys. Space Sci. \textbf{357}, 74 (2015).
	
	\bibitem{NgubelangaMaharaj20153} Ngubelanga, S.A., Maharaj, S.D.: Eur. Phys. J. Plus \textbf{130}, 211 (2015).
	
	\bibitem{NgubelangaMaharajRay2015} Ngubelanga, S.A., Maharaj, S.D., Ray, S.: Astrophys. Space Sci. \textbf{357}, 40 (2015).
	
	\bibitem{Herrera2018} Herrera, L.: Phys. Rev. D \textbf{97}, 044010 (2018).
	
	\bibitem{HerreraDiPriscoOspino2018} Herrera, L., Di Prisco, A., Ospino, J.: Phys. Rev. D \textbf{98}, 104059 (2018).
	
	\bibitem{HerreraDiPriscoOspino2020} Herrera, L., Di Prisco, A., Ospino, J.: Eur. Phys. J. C \textbf{80}, 631 (2020). 
	
	\bibitem{BogadiGovenderMoyo2022a} Bogadi, R.S., Govender, M., Moyo, S.: Eur. Phys. J. C \textbf{82}, 747 (2022).
	
	\bibitem{GovenderBarmanSharma2023} Govender, M., Barman, S., Sharma, R., Ratanpal, B.S., Govinder, K.S.: Phys. Dark Universe \textbf{42}, 101292 (2023).
	
	\bibitem{GovenderBogadiGovenderMewalal2024} Govender, M., Bogadi, R.S., Govender, W., Mewalal, N.: Astrophys. Space Sci. \textbf{369}, 25 (2024).
	
	\bibitem{MajoziMaharajGovender2026} Majozi, L.C., Maharaj, S.D., Govender, M.: Eur. Phys. J. C \textbf{86}, 529 (2026).
	
	\bibitem{MajoziGovenderMaharajGovender2025} Majozi, L.C., Govender, M., Maharaj, S.D., Govender, W.: Eur. Phys. J. C \textbf{85}, 1401 (2025).
	
	\bibitem{MajoziGovenderMaharaj2025a} Majozi, L.C., Govender, M., Maharaj, S.D.: Ann. Phys. \textbf{481}, 170156 (2025).
	
	
	
	\bibitem{TakisaMaharajMulangu2019} Takisa, P.M., Maharaj, S.D., Mulangu, C.: Pramana J. Phys. \textbf{92}, 40 (2019).
	
	\bibitem{KhuntThomasVinodkumar2023} Khunt, A.C., Thomas, V.O., Vinodkumar, P.C.: Indian J. Phys. \textbf{98}, 1571--1588 (2024).
	
	\bibitem{ZithaGovenderGovender2022}
	Zitha, V.M., Govender, W., Govender, M.:
	Eur. Phys. J. C \textbf{82}, 1029 (2022).
	
\end{thebibliography}
\end{document}